# A voltage-responsive strongly dipolar-coupled macrospin network with emergent dynamics for computing

Xinglong Ye[1,2*♪], Zhibo Zhao[3,4♪], Qian Wang[2♪], Jiangnan Li[5], Fernando Maccari[2], Ning Lu[6], Christian Dietz[2], Esmaeil Adabifiroozjaei[7], Leopoldo Molina-Luna[7], Yufeng Tian[1], Lihui Bai[1], Guodong Wang[8], Konstantin Skokov[2], Yanxue Chen[1], Shishen Yan[1*], Robert Kruk[3], Horst Hahn[3,9], Oliver Gutfleisch[2]

[1] School of Physics, Shandong University, Jinan 250100, China
[2] Institute of Materials Science, Technical University of Darmstadt, 64287 Darmstadt, Germany
[3] Institute of Nanotechnology, Kaiserstraße. 12, Karlsruhe Institute of Technology, 76131 Karlsruhe, Germany
[4] KIT-TUD-Joint Research Laboratory Nanomaterials, Technical University Darmstadt, 64287 Darmstadt, Germany
[5] Faculty of Materials Science and Engineering, Kunming University of Science and Technology, Kunming, 650031 China.
[6] School of Chemistry, Shandong University, Jinan 250100, China
[7] Advanced Electron Microscopy Division, Department of Materials and Geosciences, Technical University of Darmstadt, Peter-Grünberg-Str. 2, Darmstadt 64287, Germany
[8] State Key Lab of Crystal Materials and Institute of Crystal Materials. Shandong University, Jinan 250100, China
[9] Department of Materials Science and Engineering, University of Arizona, Tucson, AZ 85721, United States

♪ These authors contribute equally to this work. * Corresponding authors: xinglong.ye@sdu.edu.cn, shishenyan@sdu.edu.cn

**Abstract**

Emergent behavior, which arises from local interactions between simple elements, is pervasive in nature. It underlies the exceptional energy-efficient computing in our brains. However, realizing such dynamics in artificial materials, particularly under low-energy stimuli, remains a fundamental challenge. While dipole-dipole interactions are typically suppressed in magnetic storage, here we harness and amplify them to construct a strongly dipolar-coupled network of $SmCo_5$ macrospins at wafer scale, which can exhibit intrinsic interaction-driven collective dynamics in response to voltage pulses. The network combines three essential ingredients: strong dipolar coupling enabled by large single-domain macrospin; giant voltage control of coercivity over nearly 1000-fold — the largest reported to date; and a disordered network topology with frustrated Ising-like energy landscape. When stimulated by ~1 V pulses, the network enters a regime where interaction-driven magnetic behaviors emerge, including spontaneous demagnetization, greatly enhanced magnetization modulation, reversible "freeze and resume" evolution and stochastic convergence toward low-energy magnetic configurations. All these behaviors are completely absent at the single-nanomagnet level and require no external magnetic fields or high temperatures. Furthermore, by constructing micromagnetic models of the strongly dipolar-coupled macrospin networks calibrated to experiments, we show that the resulting nonlinear, high-dimensional collective dynamics, which are characteristic of strongly-interacting systems, can enable accurate chaotic Mackey–Glass prediction and multiclass drone-signal classification. Our work establishes the voltage-responsive strongly-coupled $SmCo_5$ network as a mesoscopic platform for probing emergent magnetic dynamics previously inaccessible under ambient conditions. It also suggests a fundamental distinct route towards scalable, low-voltage computing, one rooted in native physical interaction-driven collective dynamics at the network level.

**Introduction**

Emergent behavior arises when a large number of elementary components interact to produce collective, global dynamics that cannot be inferred from or reduced to the properties of individual units. Such behaviors are widespread in nature [1,2]. A paradigmatic example is the human brain, where neural firing and synaptic transmission of electrochemical potentials across the network collectively give rise to memory, learning, and cognition [3]. Remarkably, such dynamics are triggered by potential pulses as small as ~100 mV, enabling extraordinary energy efficiency. This has recently inspired intense efforts to develop neuromorphic hardware [4], physical substrates that mimic key features of individual neurons and synapses using memristors [5,6], spintronic devices [7,8,9,10,11], and phase-change devices [12]. While successful at emulating neural thresholding and synaptic plasticity at the element level, these devices often operate in piecewise fashion and must be assembled into hybrid architecture, typically with CMOS circuitry, to form functional networks. Such bottom-up engineering requires massive interconnectivity and increases architectural complexity, which leads to scalability and energy efficiency issues. More fundamentally, it remains challenging to realize genuinely emergent dynamics at the network level driven by local physical interactions.

A promising direction is to identify physical substrates in which native interactions directly govern emergent, collective dynamics. The capability of interacting units has been formalized in classical theoretical models such as Hopfield networks [13] and Boltzmann machines [14], where large numbers of binary units with pairwise couplings evolve collectively toward low-energy configurations over energy landscape. This picture is mathematically equivalent to Ising spin models at the atomic scale in magnetic materials, an analogy that originally motivated Hopfield's formulation [13,15,16]. However, atomic spin systems are difficult to probe and control. In this context, dipolar-coupled networks of single-domain nanomagnets offer an attractive mesoscopic alternative: each nanomagnet behaves as an Ising-like macrospin, which is large enough to be directly observed and accessed, and they are coupled by native dipole–dipole interactions [17,18,19]. In conventional magnetic storage, such dipole-dipole interactions are typically regarded as detrimental cross-talk. But in artificial spin ice systems, they have been exploited to enable emergent phenomena such as magnetic monopoles [20,21], crystallized magnetic charges [22] and spin-glass-like dynamics [23]. Despite these advances, two major obstacles have limited the realization of macrospin networks that evolve autonomously under their own interactions. First, the dipolar couplings are typically too weak relative to the magnetic anisotropy barrier, precluding the interaction-driven spin update under ambient conditions. Second, the control knobs typically used to activate the dynamics — high temperatures [19,22,24] or large magnetic fields [19-21] —are energetically costly and poorly matched to the low-voltage operation.

Here, we demonstrate a voltage responsive strongly dipolar-coupled network of $SmCo_5$ macrospin at wafer scale that overcomes these limitations. When stimulated by ~ 1 V pulses, the network enters a regime where dipolar interactions drive the autonomous evolution of its magnetic states. These emergent magnetic behaviors are absent at the level of individual nanomagnet, which includes spontaneous demagnetization, greatly enhanced reversible magnetization modulation and convergent, stepwise descent into thermodynamically stable states. Importantly, intermediate states along the evolution trajectory can be reversibly "frozen" and "resumed" by voltage pulses within seconds. All of these behaviors occur at room temperature without external fields. To further probe the computational capability of these emergent dynamics, we constructed micromagnetic models of strongly-coupled macrospin networks calibrated to experiments and show that its collective dynamics provide nonlinear, high-dimensional responses for temporal data processing, including accurate prediction of the chaotic Mackey–Glass time series and classification of drone radio-frequency (RF) signals. Our work tunes dipolar macrospin networks into a regime where local dipolar interaction alone can drive autonomous update of magnetic states, establishing a mesoscopic spin platform for exploring emergent magnetic phenomena under ambient conditions. It also suggests a fundamental distinct route towards scalable, energy-efficient physical substrates for information processing that exploits intrinsic physical interactions-driven collective dynamics.

**Experimental results and discussion**
**Design strategy**
Our voltage-responsive macrospin network is enabled by the integration of three key ingredients. First, we amplify dipolar coupling by using $SmCo_5$ nanomagnets as building blocks. The exceptionally high magnetocrystalline anisotropy of $SmCo_5$ stabilizes single-domain macrospin state up to sub-micrometer sizes [25], which yields large magnetic moments and correspondingly strong dipole–dipole coupling. However, this large magnetocrystalline anisotropy simultaneously leads to large coercivity, which would otherwise suppress the dipolar interaction-driven spin flipping. This dilemma is reconciled by the second key ingredients of the network: the use of the recently-developed voltage control of magnetism effect [26,27,28,29,30,31]. Here, the spin switching threshold of $SmCo_5$ nanomagnet can be tuned by nearly three orders of magnitudes from 2.1 T to 3 mT under voltage pulses, representing the largest reversible coercivity modulation reported to date [32]. This enables direct response of the network to potential pulses. When the network is tuned into low-coercivity regime, the dipolar interaction can exceed the switching energy barrier, and autonomously drive macrospin flipping and propagate collective state updates across the network. As a third ingredient, we implement a self-assembled disordered arrangement of $SmCo_5$ macrospins with distributed sizes, positions and anisotropy axes. This topology introduces competing ferromagnetic and antiferromagnetic dipolar interactions, producing a frustrated, high-dimensional Ising-like energy landscape. These three intertwined ingredients—strong dipolar coupling, giant voltage-tunability and frustrated topology—jointly lay the foundation for interaction-driven emergent dynamic behaviors.

**Figure 1A** illustrates the working principle of our voltage-responsive macrospin network. The macrospin network is immersed in aqueous $Na^+$ electrolytes, through which voltage pulses are applied. The network consists of dipolar-coupled, disordered $SmCo_5$ nanomagnets, each with magnetic moment oriented along its easy axis and represented as Ising-like macrospin, $S_i \in \{1,-1\}$. These interconnected nanomagnets evolve collectively under an effective Hamiltonian [33,34]

$$H = -\sum_{i<j} J_{ij} S_i S_j$$

where $J_{ij}$ denotes the dipolar interaction strength between nanomagnets, set by their magnetic moments, relative orientations and spatial separation. As schematically shown in Fig. 1B, C, the key control knob is the applied potentials, which can toggle the $SmCo_5$ nanomagnets between two distinct regimes: a high-coercivity state (HCS) regime at -0.4 V (left panel, Fig. 1B) and a low-coercivity state (LCS) regime at -1.2 V (right panel, Fig. 1B). In the HCS, nanomagnets act as stable memory units with $\mu_0 H_C >$ 1 T, analogous to those in conventional magnetic storage. In the LCS, however, the coercivity is substantially reduced into the mT range, so that the local dipolar fields alone can overcome the switching barrier ($\sum_j J_{ij} S_j > \mu_0 H_{C,i}$) and drive the nanomagnet toward energetically favorable state, as exemplified by the nanomagnet pairs in Fig. 1C. Consequently, when one nanomagnet flips, it perturbs its neighbors through the modified dipolar fields that further trigger subsequent cascade of asynchronous spin updates across the network (blue circles in Fig. 1A). The voltage-induced giant modulation of coercivity thus enable dipolar interaction-driven autonomous spin updates and large-scale reconfiguration of magnetic states into low-energy states.

**Structure of the macrospin network**
As the building blocks of our macrospin network, we chose $SmCo_5$ nanomagnets rather than the soft ferromagnets (e.g. permalloys) as commonly used in artificial spin ice. $SmCo_5$ is known for its exceptionally large magnetocrystalline anisotropy (~17 $MJ/m^3$), which stabilize single-domain macrospin states up to sub-micrometer sizes and yield large magnetic moments, thereby strengthening dipole–dipole coupling between neighboring nanomagnets. In parallel, we have recently shown that small electrochemical potentials can reversibly and substantially modulate the magnetic properties of $SmCo_5$ through hydrogen insertion/extraction in interstitial sites [35,36]. Under -1.2 V in aqueous electrolytes, water molecules split at the surface, and the resulting hydrogen diffuses into the $SmCo_5$ lattice (inset in Fig. 1A). This alters the crystal field environment at $Sm^{3+}$ ions and dramatically modifying the coercivity. Switching the potential back to -0.4 V drives hydrogen extraction and thus restores the magnetic properties of $SmCo_5$. This voltage-control mechanism thus allows each $SmCo_5$ nanomagnet to switch between the HCS "memory" regime and the LCS "dynamic update" regime in which dipolar interaction can drive state updates.

We built the disordered network of $SmCo_5$ nanomagnets by co-deposition of Sm and Co elements using customized magnetron sputtering. The deposition parameters, particularly influx angles, argon pressure and deposition temperatures, were tailored to enhance the shadowing effect and promote the Volmer-Weber growth that favors the formation of isolated nanoislands rather than continuous thin films. To strengthen the dipole-dipole interaction, we targeted a film thickness to ~ 150 nm while aiming for lateral nanomagnet sizes around 200 nm. Atomic force microscopy and scanning electron microscopy show that this approach produces the disordered arrays of $SmCo_5$ nanomagnets with lateral dimensions of 150–300 nm, thickness ~150 nm and nearest center-to-center distance ranging of 250–400 nm films **(Fig. 2A, Fig. S1)**. Using the saturation magnetization and the representative hexagonal nanomagnet (200×200×150 nm) with 100-150 nm edge-to-edge spacing, we estimated the dipolar fields generated by one nanomagnet to be 20-100 mT at the center of its nearest neighbor **(Fig. S2)**. This field is supposed to be sufficient to induce autonomous spin-state update when the network enters the LCS regime. X-ray diffraction reveals multiple diffraction peaks corresponding to (110), (002) and (111) planes, indicating generally random crystallographic orientation of $SmCo_5$ nanomagnets **(Fig. S3A)**. Magnetic hysteresis loops measured at different tilting angles from surface normal shows nearly the same coercivity values, further verifying the disordered distribution of magnetic easy axis across the network **(Fig. S3B)**.

We imaged the magnetic configuration of the as-prepared disordered network of $SmCo_5$ nanomagnets with magnetic force microscopy (MFM). Nanomagnets with out-of-plane magnetization components ($m_z$) pointing towards the MFM tip appear bright, while those pointing away appear dark. Fig. 2B shows that each nanomagnet exhibits a uniform contrast evidencing its single-domain state — a result consistent with the large magnetocrystalline anisotropy of SmCo$_5$. Unlike lithographically-patterned periodic arrays, our disordered network displays smoother variation of phase contrast rather than sharp bipolar contrast. This likely results from the randomly oriented easy axes and the spatial overlap of dipolar fields in the disordered network. These observations further support the picture of a disordered, strongly interacting macrospin network.

**Voltage control of single nanomagnet**
To probe how voltage pulses modulate the magnetic properties of individual SmCo$_5$ macrospin, we first measured the network's response to electrochemical potentials under large external fields of 0.2 T. This field can overwhelm local dipolar fields and allows us to approximate the behavior as that of independent nanomagnets. **Fig. 2C** shows that the pristine nanomagnet exhibits the coercivity of ~ 2.1 T under -0.4 V, consistent with the large magnetocrystalline anisotropy of SmCo$_5$. Strikingly, applying -1.2 V decreased the coercivity to 3 mT by nearly 1000-fold (**Fig. 2D**). To our knowledge, this represents among the largest voltage-induced modulation of coercivity achieved in any magnetic material reported to date [32]. This extreme tunability brings the switching threshold far below the typical dipolar fields estimated above, making it feasible for dipolar interactions to drive autonomous spin updates in the LCS regime. Meanwhile, the saturation magnetization decreases by approximately 45%, consistent with our previous measurements on micrometer-sized SmCo$_5$ powders [35].

We next evaluated the dynamical response of these voltage-controlled modulations using two experimental protocols. In the first protocol, we monitored voltage-assisted magnetization reversal (**Fig. 2E**). After saturating the network at -7 T at -0.4 V, we reserved the field to 0.2 T under which the magnetization remains negative since this field is much smaller than the coercivity (~2.1 T). However, upon applying -1.2 V pulses, the magnetization reversed within seconds, demonstrating that the coercivity reduction occurs on the same timescale (**Fig. 2E**). In the second protocol, we tracked the reversible modulation of magnetization under constant external fields of 0.2 T after saturation under 7 T, while cyclically switching the potentials between -0.4 V and -1.2 V (**Fig. 2F**). The magnetization changes by ~45% within seconds and is fully reversible over multiple cycles. These results show that each SmCo$_5$ macrospin can be reversibly toggled between a stable HCS "memory" regime and a LCS "dynamic update" regime using small voltage pulses within seconds.

**Voltage-activated emergent dynamics in the absence of magnetic fields**
We now turn to the autonomous, emergent behavior of the whole network in the absence of any external magnetic fields, where dipolar interactions dominate. The network was first initialized by applying 7 T at -0.4 V and then brought to its remanent state. A subsequent -1.2 V pulse was applied to toggle the nanomagnets from the HCS into the LCS regime. **Fig. 3A** shows that the remanent magnetization dropped sharply by ~85% within seconds, followed by a much slower decay towards a plateau. When the potential was switched back to -0.4 V, the dynamic evolution was interrupted and paused. But the magnetization only recovers to ~50% of its original value, even though the saturation magnetization of individual nanomagnets is fully restored (**Fig. 2D, F**). This partial irreversibility indicates that the network has evolved into a new low-energy configuration dictated by local dipolar coupling.

The degree of demagnetization is strongly dependent on the applied potentials. We repeated the protocol of Fig. 3A, but switched from -0.4 V to intermediate potentials between -0.6 V and -1.2 V (**Fig. 3C**). We found that the degree of demagnetization increases systematically with more negative potentials. This behavior is consistent with the fact that more negative potentials induce larger hydrogen concentration within the $SmCo_5$ lattice, leading to stronger reduction in magnetocrystalline anisotropy and thus facilitating more extensive dipolar-driven macrospin flipping. To further quantify the role of dipolar fields, we measured the voltage-induced demagnetization at -1.2 V under different external magnetic fields. **Fig. S4** shows that only for fields above ~50 mT, the initial magnetization can be fully restored. This further reveals the critical role of dipolar interaction: only external fields over 50 mT are sufficient to counteract the local dipolar fields and suppress collective reorientation, consistent with our earlier estimates of dipolar strength (20–100 mT).

Once the network has been driven into a voltage-demagnetized state, subsequent potential cycling produces a dramatically enhanced reversible modulation of magnetization. **Fig. 3B** shows that switching the potential between -1.2 V and -0.4 V yields a reversible change of remanent magnetization ($M_{-0.4 V}/M_{-1.2 V}$) as large as ~450%, approximately twice the modulation at the single-nanomagnet level (**Fig. 2F**). Thus, such amplification cannot arise from intrinsic voltage-induced changes in individual nanomagnets. Instead, it again verifies cooperative macrospin reorientation driven by local dipolar interactions across the network in the LCS regime. This macroscopic amplification is another hallmark of interaction-dominated systems, indicating that the network behaves as a strongly-coupled collective entity rather than a sum of independent nanomagnets.

We further directly visualized the voltage-triggered demagnetization by performing correlated AFM and MFM imaging on the same region of the network across three distinct magnetic states: the thermally-demagnetized (as-deposited) state, the remanent state after initialization, and the voltage-demagnetized state. AFM images confirm that the surface morphology remained nearly the same throughout, verifying the non-destructive nature of voltage control and consistent with the earlier observed reversibility (**Fig. S5**). In stark contrast, the MFM images reveal pronounced changes in their magnetic configurations (**Fig. 3D**). In the as-deposited state, individual nanomagnets display random bright and dark features, corresponding to upward and downward $m_z$ components, indicative of thermally-disordered macrospin orientation. After initialization at -0.4 V under 7 T, the remanent state displays uniform bright contrast across the network, reflecting a fully magnetized configuration with all $m_z$ pointing upward (middle panel, **Fig. 3D**). Upon applying -1.2 V, however, dark features reappear across the network, directly evidencing voltage-induced spontaneous macrospin flipping (right panel, Fig. 3D). To highlight the reconfigured spins, we subtracted the MFM images of the demagnetized and voltage-modulated states from the fully magnetized image (**Fig. S5**). The resulting differential maps reveal widespread spin flipping triggered by small voltages, directly visualizing interaction-driven macrospin flipping without the need of magnetic fields or high temperatures.

**Stochastic spins and energy descent in tilted energy landscapes**
To probe how the network explores its energy landscape under weak bias, we introduce a small external field of 3 mT that is far below the dipolar fields (~ 50 mT) so as not to overwhelm the dipolar interactions (~50 mT, **Fig. S2, S4**). After initialization at -7 T at -0.4 V, we applied the bias field and then perturbed the network with potential pulses of -1.2 V. As shown in **Fig. 4A**, the magnetization evolved continuously: it decreases gradually, crosses zero and eventually approaches a plateau over one hour—roughly two to three orders of magnitude slower than the rapid voltage-induced changes observed at the single-nanomagnet level. The same behavior was reproduced for the networks initialized with opposite magnetization polarity **(Fig. S6)**. The prolonged, slow relaxation suggests that spin flips in this regime are thermally activated and stochastic. Consequently, the macrospins converge towards a thermodynamically stable configuration determined by the interplay between bias fields and dipolar coupling. To isolate the intrinsic time scales of collective dynamics without external magnetic fields, we also examined voltage-triggered self-demagnetization under zero fields by maintaining -1.2 V continuously without interruption (unlike the interrupted protocol in **Fig. 3A**). **Fig. 4B** shows that the magnetization gradually decreases and plateaus over extended time scale. When plotted on a double-logarithmic scale, the relaxation follows a power-law decay with exponent ~0.22, reminiscent of canonical spin-glass systems [37] and dipolar-coupled nanoparticle ensembles at high temperatures [38,39].

Crucially, intermediate states along this convergence trajectory can be reversibly "frozen" and "resumed" within seconds by switching between -0.4 V and -1.2 V. When the potential is returned to -0.4 V, the instantaneous configuration is locked in and memorized; switching back to -1.2 V reactivates the dynamics and allows further stepwise descent towards lower-energy states. This voltage-controlled "pause-and-play" functionality demonstrates the system's ability to explore complex energy landscapes in a controlled yet autonomous manner. Overall, these measurements establish that the SmCo$_5$ macrospin network acts as a physical substrate that stochastically explores its energy landscape under small voltage perturbations and eolves toward low-energy configurations driven by local dipolar interaction.

**Using the interaction-driven network collective dynamics for processing temporal information**
The emergent behaviors demonstrated above show that the strongly dipolar-coupled SmCo$_5$ macrospin network possesses essential properties desirable for processing temporal information. In response to voltage pulses, the network evolves nonlinearly, exhibits high effective dimensionality due to the extremely large number of interacting macrospins ($10^7$–$10^8$ per mm$^2$), and progressively converges toward voltage-dependent configurations with fading memory. These are key attributes required for the physical reservoir computing [40,41]. However, directly accessing the full internal state of all these nanomagnets in real time is currently not feasible. To explore potential computational capabilities, we constructed simplified micromagnetic models of a disordered, strongly dipolar-coupled SmCo$_5$ macrospin network. We used the network sizes of 440 dipolar-coupled macrospins, providing the tractable, nontrivial and high-dimensional state space for testing temporal tasks. The goal is to capture the essential ingredients of the physical macrospin network—strong dipolar coupling, voltage-tunable macrospin stability, and the disordered topology—and to assess whether the resulting dynamics can support standard RC tasks. In the simulations, the effect of voltage pulses is mapped to effective magnetocrystalline anisotropy ($K_U$) between $10^4$ and $10^6$ J m$^{-3}$, calibrated to reproduce the measured coercivity modulation. This mapping is only approximate, but it retains the key feature that lowering $K_U$ makes individual macrospins more susceptible to dipolar-driven reconfiguration, in analogy to the voltage-controlled coercivity reduction in the experiments.

We first assessed the temporal prediction capability of this model reservoir to predict the chaotic Mackey–Glass (MG) time series. A normalized MG sequence $x(t_i)$ is converted into a sequence of effective anisotropy values $K_U(t_i)$, which drives the strongly-coupled macrospin network to reconfigure its magnetic state (**Fig. S7A**). For each time step, the macrospins evolve collectively under the corresponding $K_U(t_i)$, and their out-of-plane components $m_{z,j}(t_i)$ served as the reservoir state. **Fig. 5B** and **Fig. S7C** show that the temporal evolution of the global magnetization exhibits nonlinear, reversible transitions between high- and low-magnetization regimes for $K_U = 10^4$ and $10^6$ J m$^{-3}$, consistent with the experimental voltage-controlled modulation of remanent magnetization (**Fig. 3B**). This confirms that the constructed macrospin model captures the essential dipolar-driven collective dynamics of the physical SmCo$_5$ network. Then, we used the resulting 100-dimensional state vector, each containing 3-4 macrospins, to provide spatial high-dimensional temporal encoding of the input sequence (**Fig. S7B**), which are trained offline to predict future MG amplitudes $x(t+\tau_p)$ with $\tau_p$ as horizontal steps. **Fig. 5C** shows that, the trained reservoir reproduces the chaotic MG trajectory with high accuracy for $\tau_p=1$, and maintains excellent performance at longer horizons, achieving mean-square errors (MSE) of ~$2.37\times10^{-3}$ and $5.49\times10^{-3}$ for $\tau_p = 10$ and 20 steps, respectively. This performance is better than that obtained using artificial spin-vortex ice driven by external magnetic fields [42] and using magnetic skyrmion dynamics [43]. In contrast, a baseline model that attempts to map the raw MG input directly to $x(t+\tau_p)$ fails completely by $\tau_p = 10$.

We next consider a real-world RF signal recognition task—drone classification. For this demonstration, we used a public 10-class RF dataset comprising nine drone or remote-controller classes and one Wi-Fi router class [44]. The 256×256 time–frequency spectrogram were averaged along the time axis to obtain a 256-dimensional spectral-amplitude vector. These features are then converted and mapped into $K_U(t_i)$ sequence and applied to the simulated macrospin network reservoir. As shown in **Fig. 5D**, the obtained confusion matrix reveals the maximum recognition rate of 98% (94.03 ± 1.88% on average) on the test set, comparable to the performance of multilayer spintronic neural network driven by large currents after multiple iterations [45]. These two demonstrations show that the strongly dipolar-coupled SmCo$_5$ macrospin network, when perturbed by time-dependent $K_U$ inputs, can generate rich spatiotemporal dynamics with high dimensionality suitable for computing.

While the above demonstrations rely on simulated access to all macrospin states, they highlight how the experimentally observed voltage-responsive collective dynamics could, in principle, be probed experimentally for temporal information processing. In practice, the internal reservoir state would not have to be read out at the single-macrospin level. Instead, anomalous Hall voltages or magnetoresistance [46] measured at multiple electrodes patterned across the wafer-scale network could already provide sufficient large number of independent output, each integrating the collective response of thousands of nanomagnets. Moreover, because macrospins interact through intrinsic dipolar coupling and update in parallel under voltage pulses, the associated energy consumption is determined primarily by the electrochemical charging of hydrogen atoms. Using typical voltage pulses of 1 V and charging current of ~0.1 mA for ~1 s across 10 mm² samples, this corresponds to an energy scale of ~$10^{-4}$ J per global update, or roughly sub-pJ per macrospin when distributed across ~$10^8$ interacting macrospins. Although approximate, this underscores the potential for low-energy operation arising from the intrinsic parallel macrospin update and dipolar-driven collective dynamics of the network.

**Conclusion**

We have established a wafer-scale, strongly dipolar-coupled network of $SmCo_5$ macrospins by amplifying the dipolar interactions, which can respond to small voltage pulses and is capable of autonomous, native interaction-driven collective dynamics towards low-energy states. This network integrates three core ingredients: the strong dipole-dipole interaction, voltage-controlled giant modulation of coercivity, and a frustrated Ising-like energy landscape. When perturbed by small voltages of only ~ 1V, the network exhibits interaction-driven behaviors absent at the single-nanomagnet level, including self-demagnetization, large-scale coordinated reconfiguration, and stochastic, convergent evolution towards thermodynamic stable states. These dynamics can be paused and resumed by small voltages, enabling the transitions between stable memory configurations and the dynamic update state. Using micromagnetic simulation of such a strongly dipolar-coupled network, we show the resulting high-dimensional collective dynamics can process temporal data, achieving accurate chaotic Mackey-Glass time-series forecasting and classification of real-world drone signals. Our work establishes the voltage-responsive, strongly dipolar-coupled macrospin network as a nanoscale spin system, which can be directly probed and controlled at ambient conditions, for exploring emergent magnetic phenomena in previously inaccessible regimes. Our works also suggests an alternative route for developing scalable, energy-efficient physical substrate for information processing, one rooted not in emulating individual neurons or synapses but in native interaction-driven emergent dynamics.

**Figures and figure captions are shown in the following pages**

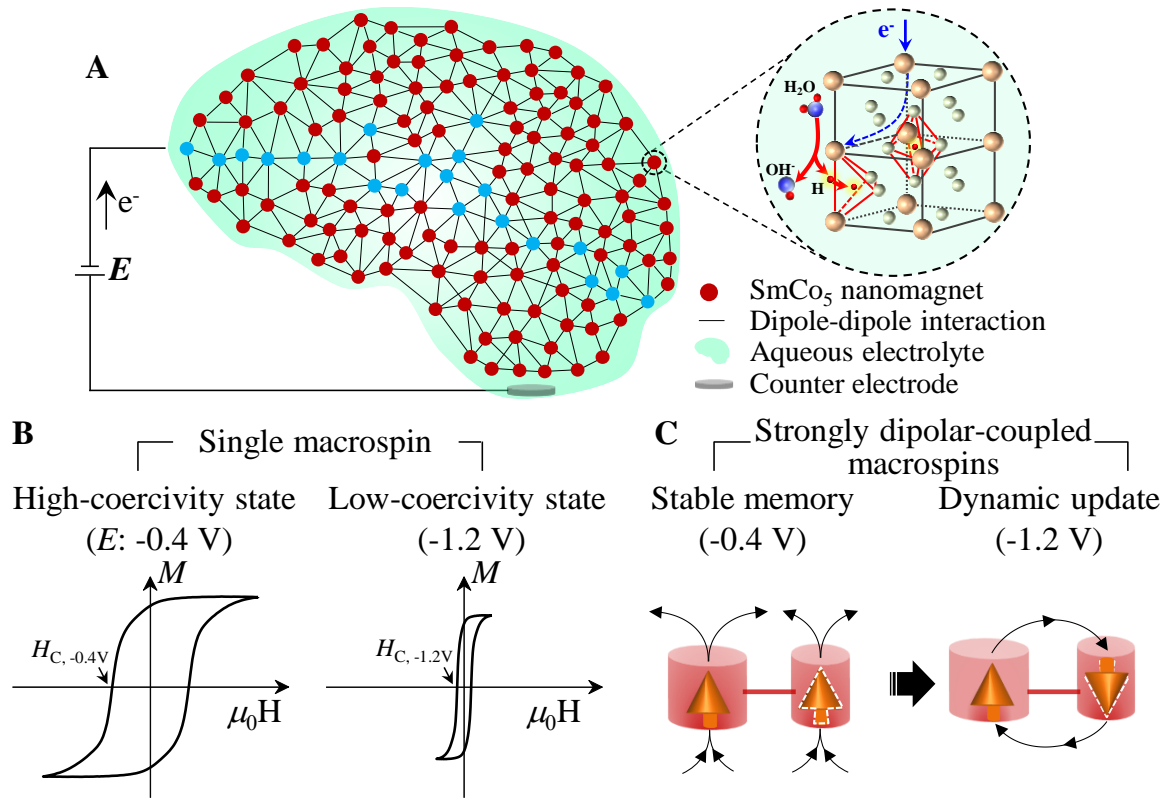

**Fig. 1 Design strategy of the voltage-responsive macrospin network**
(A) Schematic of the self-assembled, disordered $SmCo_5$ nanomagnets network immersed in aqueous electrolytes. Each $SmCo_5$ nanomagnet behaves as a single-domain Ising-like macrospin and is strongly coupled to its neighbors through dipole-dipole interactions. The side panel illustrates the mechanism underlying voltage control of coercivity in $SmCo_5$ nanomagnets through electrochemically-driven insertion of hydrogen atoms into interstitial sites of crystal lattice. (B) Voltage control of single nanomagnet. Individual nanomagnets can be toggled between two distinct states: high-coercivity state (HCS) at -0.4 V and the low-coercivity state (LCS) at -1.2 V with giant modulation of coercivity. (C) Dipolar interaction-driven collective dynamics. In the HCS, $SmCo_5$ nanomagnets function as stable memory elements, as in conventional magnetic storage ($H_{C,-0.4 V} \gg H_{dipole}$). When the network enters the LCS regime, the coercivity is reduced far below the local dipolar fields, so that dipolar interaction can drive macrospin flipping and reorientation towards low-energy configurations. This is exemplified by spontaneous switching from parallel to antiparallel alignment in a coupled nanomagnet pair.

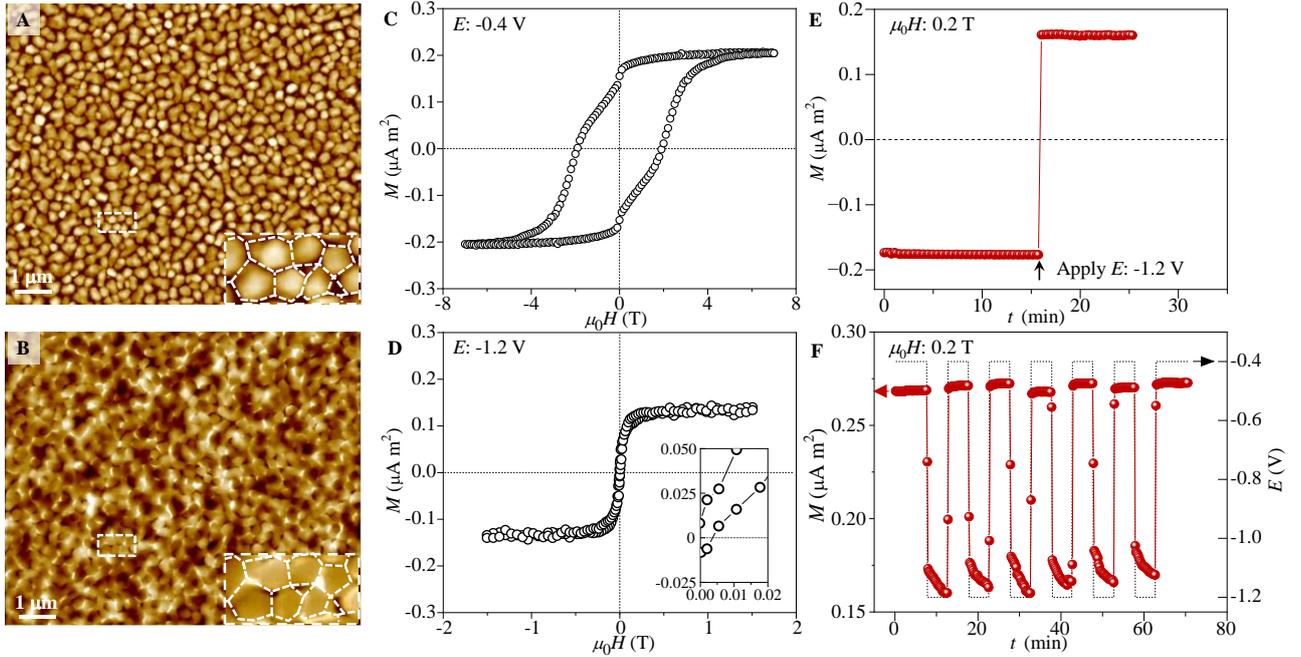

**Fig. 2 Structure of the network and voltage control of single nanomagnet**
(A) AFM image of the self-assembled, disordered network of SmCo$_5$ nanomagnets. (B) Corresponding MFM image showing single-domain macrospin state and local coupling to 4-6 nearest neighbors. The nanomagnets have large lateral dimensions of 150 to 350 nm. (C, D) In-situ magnetic hysteresis loops measured at applied potentials of -0.4 V and -1.2 V, respectively. The coercivity is reversibly tuned from 2.1 T to 3 mT —an unprecedented ~1000-fold modulation, accompanied by ~40% reduction in saturation magnetization. The inset in (D) highlights the extremely low coercivity at -1.2 V. (E) Voltage-assisted magnetization reversal under constant magnetic fields of 0.2 T. The network was firstly magnetized by applying -7 T under -0.4 V, and then the field was reversed to 0.2 T, under which the magnetization remains negative. Applying -1.2 V pulses induced rapid magnetization reversal within several seconds. (F) Voltage-controlled reversible modulation of saturation magnetization under 0.2 T by alternating -0.4 V and -1.2 V, demonstrating rapid and reversible tunability. Note that the hysteresis loops shown in (C) and (D) were acquired *in situ* for the same sample, while in (E) and (F) two different samples were used.

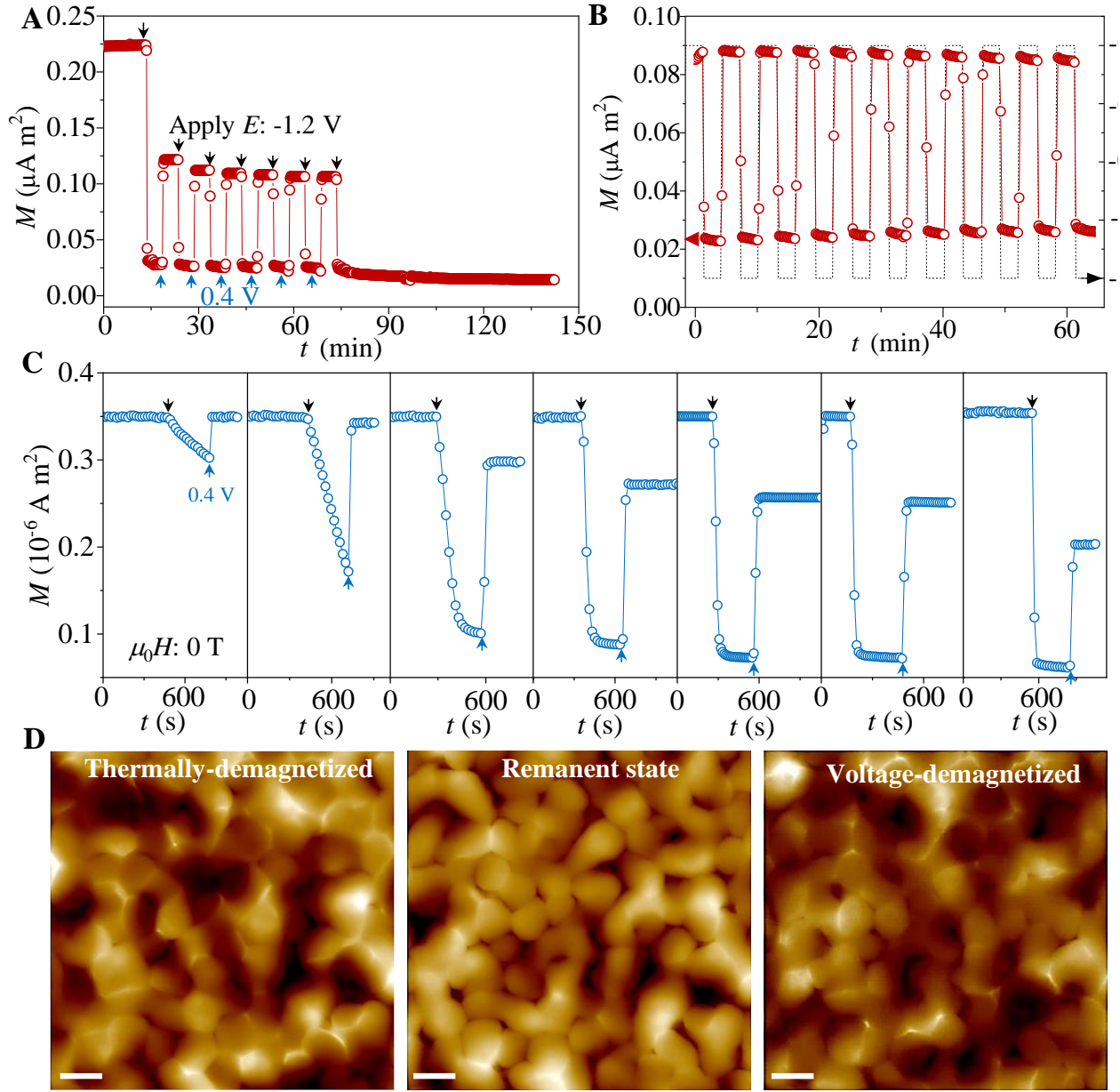

**Fig. 3 Voltage-activated emergent dynamics without external magnetic fields**
(A) Voltage-triggered autonomous demagnetization. After preparing remanent states by applying 7 T under -0.4 V, the subsequent application of -1.2 V pulse induced a rapid decrease in magnetic moment values. Switching the voltage back to -0.4 V restores only ~50% of the original moment, indicating that the network has evolved into a new, dipolar interaction-dominated low-energy configuration. (B) Significantly enhanced, reversible modulation of remanent magnetization in the voltage-demagnetized state. Alternating -1.2 V and -0.4 V potentials modulated magnetic moments by ~450% ($M_{-0.4V}/M_{-1.2V}$), nearly twice the individual-nanomagnet modulation in Fig. 2C-F. Black and light blue arrows mark the timing of each voltage pulse. (C) Dependence of the demagnetization on the applied potentials from -0.6 V to -1.2 V, showing progressively stronger demagnetization with more negative potentials. (D) MFM images of the same region in three states: thermally-demagnetized as-grown state (left), remanent state after full magnetization at -0.4 V under 7 T (middle), and voltage-demagnetized state at -1.2 V (right). Scale bar is 200 nm. The contrast evolution—from random to uniform to mixed—directly visualizes large-scale spontaneous macrospin flipping driven by dipolar interactions.

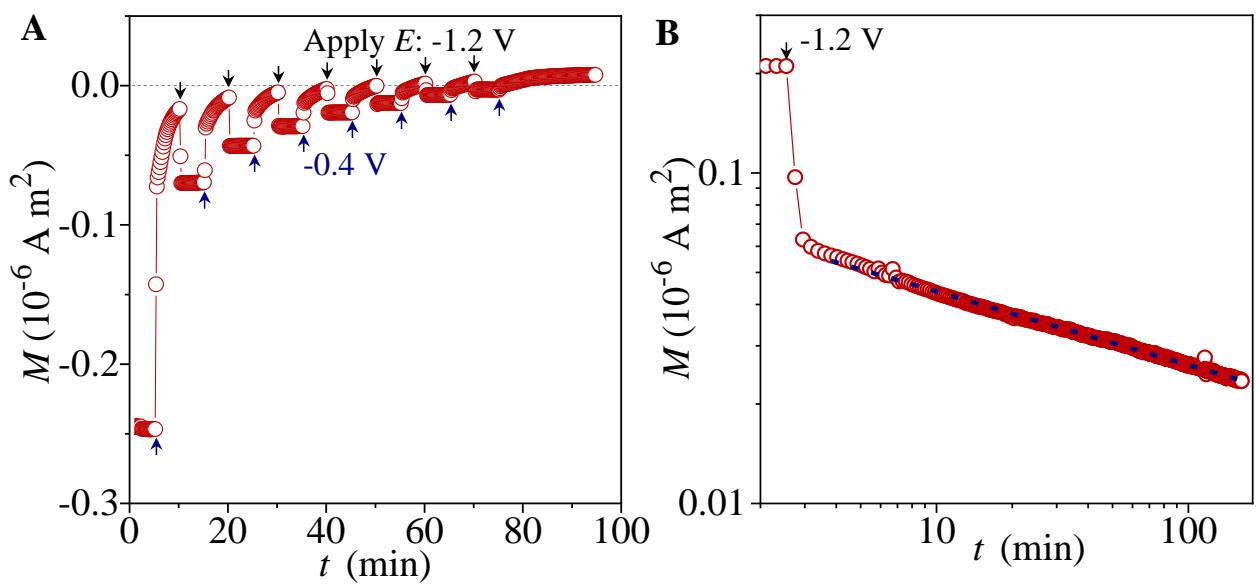

**Fig. 4 Stochastic, convergent evolution towards thermally stable states**
(A) Dynamic evolution of magnetization under small bias fields of 3 mT, with the potentials alternated between -1.2 V and -0.4 V. Applying -1.2 V triggers a slow, convergent flow of magnetization over time scales exceeding 1 hour, nearly three orders of magnitude longer than the sub-10 s response of individual nanomagnet in Fig. 2C-F. Intermediate states along this trajectory can be rapidly "frozen" into stable memory regime and "resumed" by applying -1.2 V, enabling stepwise exploration of the energy landscape. (B) Voltage-triggered self-demagnetization plotted on a double-logarithmic scale, revealing a power-law evolution of magnetization. This plot corresponds to the uninterrupted protocol shown Fig. 3A (i.e., without -0.4 V pauses).

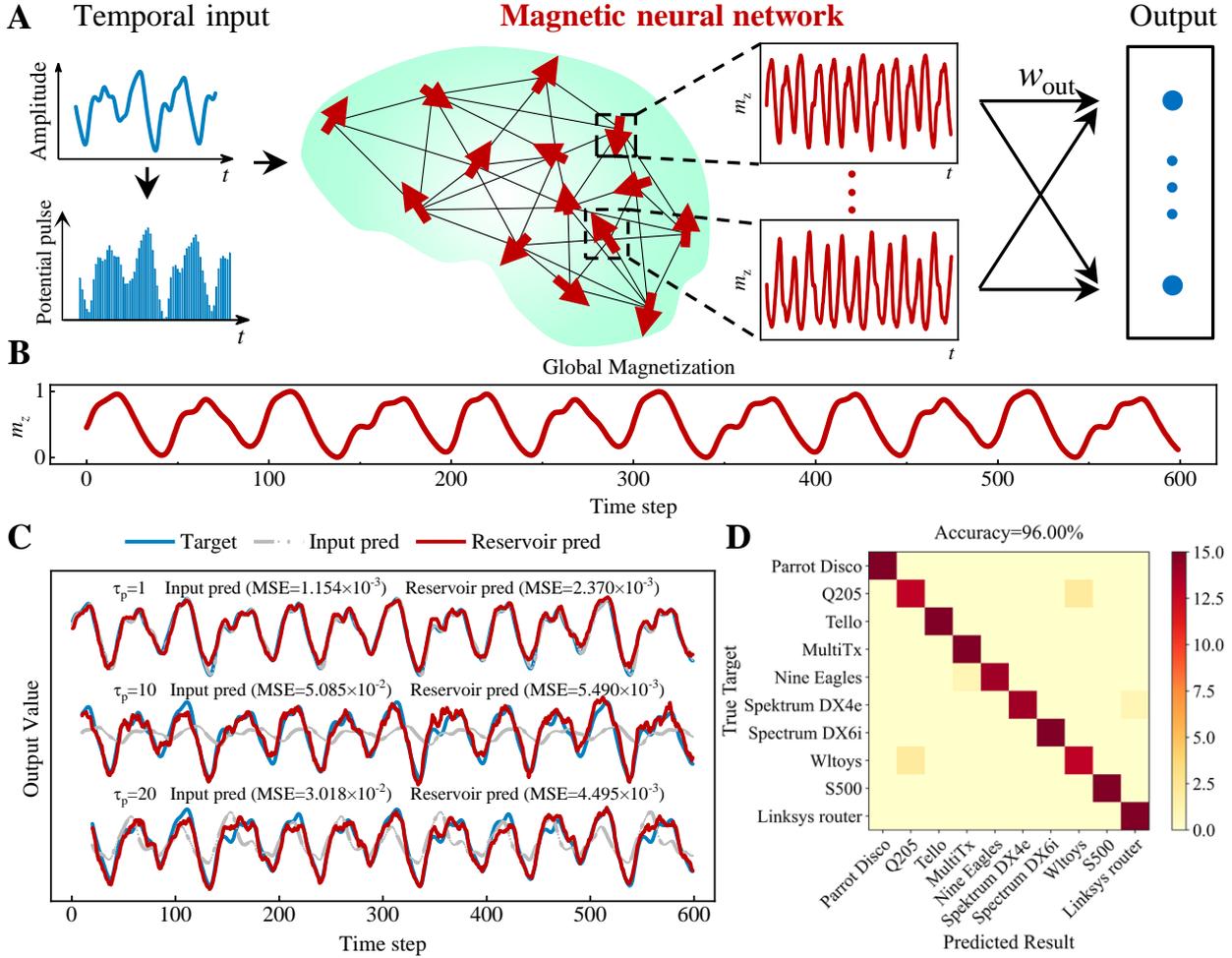

**Fig. 5 Using interaction-driven network dynamics for prediction and classification tasks**
(A) Schematic illustration of the physical reservoir computing scheme implemented through micromagnetic simulation of a strongly dipolar-coupled network calibrated to experiments. Temporal input signals are mapped to sequences of potential pulses between -0.4 V and -1.2 V, which are converted to time-dependent magnetocrystalline anisotropy values ($10^6$–$10^4$ J/m$^3$). The simulated network in the demagnetized state evolves collectively under these anisotropy changes, and the out-of-plane magnetization $m_{z,j}(t)$ of individual macrospin form the high-dimensional reservoir state used for ridge-regression training. (B) Simulated dynamic evolution of the global remanent magnetization ($m_z$) under pulsed change of effective anisotropy, demonstrating dipolar interaction-driven coordinated macrospin reorientation consistent with the experimental results in Fig. 3B. (C) Forecasting of the chaotic Mackey–Glass time-series. The trained reservoir accurately predicts future MG amplitudes for horizontal steps ($\tau_p$) of 1, 10 and 20, substantially outperforming the baseline linear model applied directly to raw data. (D) The confusion matrix of drone RF signal recognition, achieving the maximum recognition rate of 98% (94.03±1.88% on average).

**Acknowledgements**

XY conceptualized the work and designed the experiments. ZZ prepared $SmCo_5$ thin films and conducted XRD measurements. XY performed in-situ magnetometry measurements. FM, CD and XY conducted AFM and MFM measurements. JL conducted micromagnetic simulations and QW performed reservoir computing tests. EA performed TEM, NL analyzed MFM images and GW performed SEM characterization. XY, YT, LB and YC together analyzed the data, and XY and SY wrote the draft of manuscript and all authors contributed to the discussion, and the revision of the manuscript. XY acknowledge financial support by the Deutsche Forschungsgemeinschaft (DFG) under grant number 528530757 and NSFC Excellent Young Scientists Program. ZZ, RK and HH acknowledge the support from the European Union's Horizon 2020 research and innovation program under the Marie Skłodowska-Curie grant agreement No.861145.

Supplementary materials for

# A voltage-responsive strongly dipolar-coupled macrospin network with emergent dynamics for computing

Xinglong Ye[1,2*♪], Zhibo Zhao[3,4♪], Qian Wang[2♪], Jiangnan Li[5], Fernando Maccari[2], Ning Lu[6], Christian Dietz[2], Esmaeil Adabifiroozjaei[7], Leopoldo Molina-Luna[7], Yufeng Tian[1], Lihui Bai[1], Guodong Wang[8], Konstantin Skokov[2], Yanxue Chen[1], Shishen Yan[1*], Robert Kruk[3], Horst Hahn[3,9], Oliver Gutfleisch[2]

[1] School of Physics, Shandong University, Jinan 250100, China
[2] Institute of Materials Science, Technical University of Darmstadt, 64287 Darmstadt, Germany
[3] Institute of Nanotechnology, Kaiserstraße. 12, Karlsruhe Institute of Technology, 76131 Karlsruhe, Germany
[4] KIT-TUD-Joint Research Laboratory Nanomaterials, Technical University Darmstadt, 64287 Darmstadt, Germany
[5] Faculty of Materials Science and Engineering, Kunming University of Science and Technology, Kunming, 650031 China.
[6] School of Chemistry, Shandong University, Jinan 250100, China
[7] Advanced Electron Microscopy Division, Department of Materials and Geosciences, Technical University of Darmstadt, Peter-Grünberg-Str. 2, Darmstadt 64287, Germany
[8] State Key Lab of Crystal Materials and Institute of Crystal Materials. Shandong University, Jinan 250100, China
[9] Department of Materials Science and Engineering, University of Arizona, Tucson, AZ 85721, United States

♪ These authors contribute equally to this work.
* Corresponding authors: xinglong.ye@sdu.edu.cn, shishenyan@sdu.edu.cn

**This file includes:**
Fig. S1 to S7

**Fabrication of wafer-scale, disordered SmCo$_5$ nanomagnet network**

We fabricated the self-organized SmCo$_5$ nanomagnet network using DC magnetron co-sputtering of Sm and Co targets onto thermally-oxidized silicon substrates at a substrate temperature of 650 °C. A 20 nm Cr buffer layer was used to improve adhesion between the SmCo$_5$ layer and the substrate. The base pressure of the sputtering chamber was maintained below 2×10$^{-9}$ Torr, and the deposition process was conducted under an Ar atmosphere at 3 mTorr. The desired stoichiometry of SmCo$_5$ was achieved by precisely adjusting the sputtering power of the Sm (9 W) and Co (50 W) targets. To induce Volmer–Weber (island) growth, we carefully optimized the sputtering geometry. The incident flux angle was increased to 37.5°, and the target-to-substrate distance was extended to approximately 300 mm. These modifications enhanced the geometrical shadowing effect, which together promotes localized nucleation and island growth. This results in the growth of spatially-separated SmCo$_5$ nanomagnets rather than continuous films, a key feature necessary for the dipolar-coupled network structure. The lateral dimensions and inter-nanomagnet spacing were engineered by adjusting the deposition rate and the deposition time, which in turn sets the dipole-dipole interaction strength. After growth of the SmCo$_5$ network, a 10 nm Pd capping layer was deposited at room temperature. This capping layer prevents oxidation of the SmCo$_5$ nanomagnets and facilitates efficient hydrogen transport during the electrochemical control of hydrogen insertion/extraction within the SmCo$_5$ lattice.

**Structural and microstructure characterization**

We employed multi-technique characterization approaches to analyze the morphology, composition, and magnetic domain structure of the self-organized SmCo$_5$ nanomagnet network. The surface morphology and elemental composition of the SmCo$_5$ nanomagnet network were examined using Scanning Electron Microscopy (SEM, ZEISS G560) equipped with Energy-Dispersive X-ray Spectroscopy (EDS). EDS mapping provided spatially resolved information on the distribution and uniformity of Sm and Co across the nanomagnet network. The crystal structure of the SmCo$_5$ nanomagnet network was characterized by X-ray Diffraction (XRD) with a Cu Kα radiation source. Cross-section transmission electron microscopy (TEM) was further used to characterize the thickness and morphology of the nanomagnet network. TEM samples were prepared by standard focused-ion-beam (FIB) lift-out and thinning procedures using FIB/SEM systems (FEI Strata 400 and Zeiss Auriga 60). Surface topography and magnetic domain structure were mapped by atomic force microscopy (AFM, Bruker Dimension Icon) using high-coercivity cantilevers coated with Co-Pt (ASYMFMHC-R2, Asylum Research). MFM imaging was performed in dual-pass mode: the first pass operated in tapping mode to acquire high-resolution surface topography, followed by a second pass at a constant lift height of 30 nm to record magnetic signals. To enhance the phase contrast, we magnetized the network using out-of-plane magnetic fields prior to all MFM measurements.

*In situ* **magnetometry measurements**

To evaluate the magnetic response of the self-organized SmCo$_5$ nanomagnet network under voltage stimuli, we employed a custom-designed three-electrode electrochemical cell integrated into a superconducting quantum interference device (SQUID, Quantum Design MPMS 3). This setup enabled real-time, *in situ* measurements of the magnetic properties during the application of voltage pulses. The electrochemical cell consisted of a SmCo$_5$ nanomagnet network film as the working electrode, Pt foils as the counter electrode, and a pseudo-Ag/AgCl reference electrode, all controlled by an Autolab PGSTAT302N potentiostat. This pseudo-Ag/AgCl reference electrode was calibrated against a Hg/HgO (1 M KOH) reference and found to be 0.300 ± 0.002 V more positive than Hg/HgO. The electrolyte was 0.1 M KOH aqueous solution prepared using ultrapure water (resistivity ~18.2 MΩ·cm).

**Micromagnetic Model**

We simulated the SmCo$_5$ nanomagnet network on a regular finite-difference grid of 500 × 500 × 18 cells with 3 nm cubic spacing, corresponding to a physical size of 1.5 μm × 1.5 μm × 54 nm. The network was portioned into 440 nanomagnets. Each nanomagnet is assigned first-order uniaxial anisotropy whose easy axis lies in the x-z plane and is biased toward z axis, mimicking the experimentally observed distribution of easy axes. Material parameters were set to $M_s$=7.8×10$^5$ A/m and exchange stiffness A=1.0×10$^{-11}$ J/m. The uniaxial anisotropy constant $K_u$ was treated as time-varying according to the temporal input and drives the emergent dynamics of the network. In the model, the experimental voltage pulses are represented as a change in $K_u$, with the voltage range of -0.4 V to -1.2 V mapped to an anisotropy range $K_u \in [10^5, 10^7]$ J/m$^3$. A normalized scalar input s ∈ [0,1] is converted to magnetic anisotropy via

$$K(s)_u = K_{u,min} + (K_{u,max} - K_{u,min})s$$

To read out the spatiotemporal magnetic response, we sampled 100 uniformly distributed spatial blocks. Each region covered 30×30 computational cells, corresponding to a physical area of 90 nm × 90 nm. For each input $K_u$, the film was evolved for 40 relaxation steps using an energy-descent update operator, and only the final state of all 100 regions was recorded. Stacking these states over time yields a 100×T spatiotemporal response matrix, where T is the length of the input sequence.

**Mackey-Glass Task**

Time-series inputs are generated from the Mackey–Glass (MG) system:

$$\frac{dx(t)}{dt} = \beta \frac{x(t-\tau)}{1 + [x(t-\tau)]^n} - \gamma x(t).$$

By setting parameters τ=17, β=0.2, γ=0.1, n=10, and the initial value $x_0$=1.2, the system corresponds to chaotic dynamics (chaotic behaviors can be observed for τ > 16.8). The Mackey–Glass differential equation was discretized using a fixed time step of Δt = 1, which defines the sampling interval of the generated MG sequence. The resulting sequence x(t) is subsequently normalized to s ∈ [0,1] and then mapped nonlinearly to the anisotropy range as stated above ($K_u$ as $K_u$=10$^{4+2s}$) J/m$^3$. For each input $K_u$, the network relaxed for 40 steps as described above, and the final out-of-plane magnetization of each of the 100 regions was recorded, yielding a 100-dimensional reservoir state. Stacking these states over time gives a T×100 matrix, where each row corresponds to one time index and each column to one block. For MG forecasting, we construct one supervised example per time index t by pairing the 100-dimensional reservoir state with a future target $x(t+\tau_p)$, where $\tau_p$ is the prediction horizon. We consider $\tau_p$=1, 10, 20, and fix the regularization to α=10$^{-3}$. The linear readout weights W are obtained by ridge regression

$$W = (X^\top X + \alpha I)^{-1} X^\top y.$$

We use 2,000 MG points for training and approximately 600 MG points for testing ((the exact number of test points depends on $\tau_p$, since a valid target requires the availability of $x(t+\tau_p)$). Consequently, the dimensions of input training matrix $X$, target training matrix $y$, and weight matrix $W$ are 2000×100, 1×2000, and 1×100, respectively.

**Drone Classification**

For drone recognition we use a public 10-class radio-frequency (RF) dataset containing 1,000 samples in total, including nine drone/remote-controller classes and one Wi-Fi router. The RF waveforms were segmented and transformed using the short-time Fourier transform (STFT) to produce 256×256 time–frequency spectrograms as samples, which is made public by Basak et al. We averaged spectrograms over time frames to obtain discrete Fourier transforms with an averaged amplitude. The 256 amplitudes were then normalized to the range of s $\epsilon$ [0,1] and then nonlinearly mapped into anisotropy sequence as $K_u=10^{4+2s}$ J/m³. The different anisotropy amplitudes were fed into the reservoir system over time (per $K_u$ compute 40 steps), yielding a dynamic response matrix of size 256×100, where the 100 spatial measurements correspond to 100 uniformly distributed blocks as described above. Flattening this matrix yields a 25600-dimensional feature vector per sample. We trained a linear classifier using ridge regression, randomly selecting 15 samples per class for testing and using the remainder for training. The training set contains 850 samples, while the test set contains 150 samples. Therefore, the 850 training dataset composes a 850×25600 dimensional matrix, and the corresponding training target is 10×850 dimensional. The classifier weight matrix W is then solved in closed form using ridge regression:

$$W = (X_{train}^\top X_{train} + \alpha I)^{-1} X_{train}^\top y_{train}$$

This expression gives the optimal linear mapping from the reservoir feature space to the ten output scores. The size of resulting weight matrix $W$ is 10×25600, which contains one weight vector for each class. The readout layer is a multi-class linear classifier trained via ridge regression. The target labels are converted into one-hot, where each row contains a single "1" marking the correct class and nine "0" entries for the other classes. This representation allows the regression model to learn 10 dependent output channels corresponding to the drone categories. During testing, a sample is classified by multiplying its feature vector with $W$. The class with the largest output score is selected as the prediction:

$$y_{test} = argmax(WX_{test})$$

To reduce the influence of a particular random split, we repeat the random stratified partitioning and ridge-regression training R=100 times, and the mean and standard deviation of the test accuracy over these 100 runs are 94.03% and ±1.88%, where the standard deviation is the population standard deviation:

$$\sigma = \sqrt{\frac{1}{R}\sum_{r=1}^{R}(a_r - \bar{a})^2}$$

where $a_r$ is the test accuracy obtained in run r and $\bar{a}$ their mean.

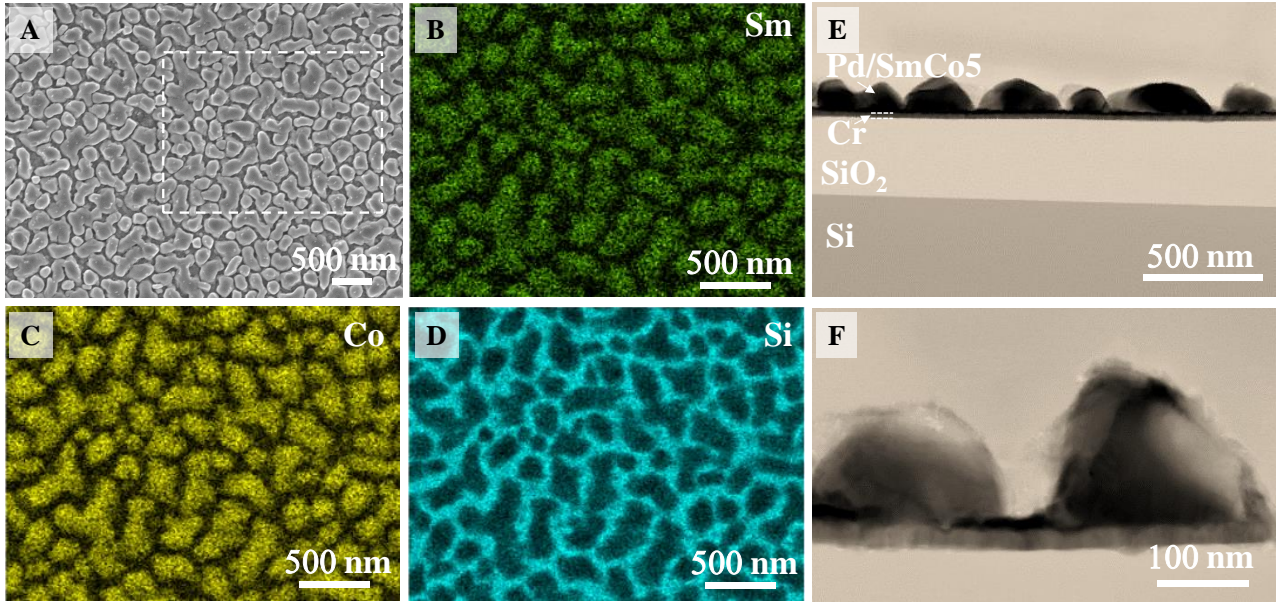

**Fig. S1 Microstructure characterization of the SmCo$_5$ nanomagnet network**
(A) SEM micrograph of the self-assembled SmCo$_5$ nanomagnet network, showing spatially-isolated SmCo$_5$ nanomagnets with distributed sizes and positions. (B-D) Corresponding element distribution of Sm, Co and Si, verifying the network topology of the SmCo$_5$ nanomagnets. (E, F) Cross-sectional bright-field TEM image revealing the film stack configuration of SmCo$_5$/Cr/SiO$_2$/Si with an average SmCo$_5$ thickness of ~150 nm.

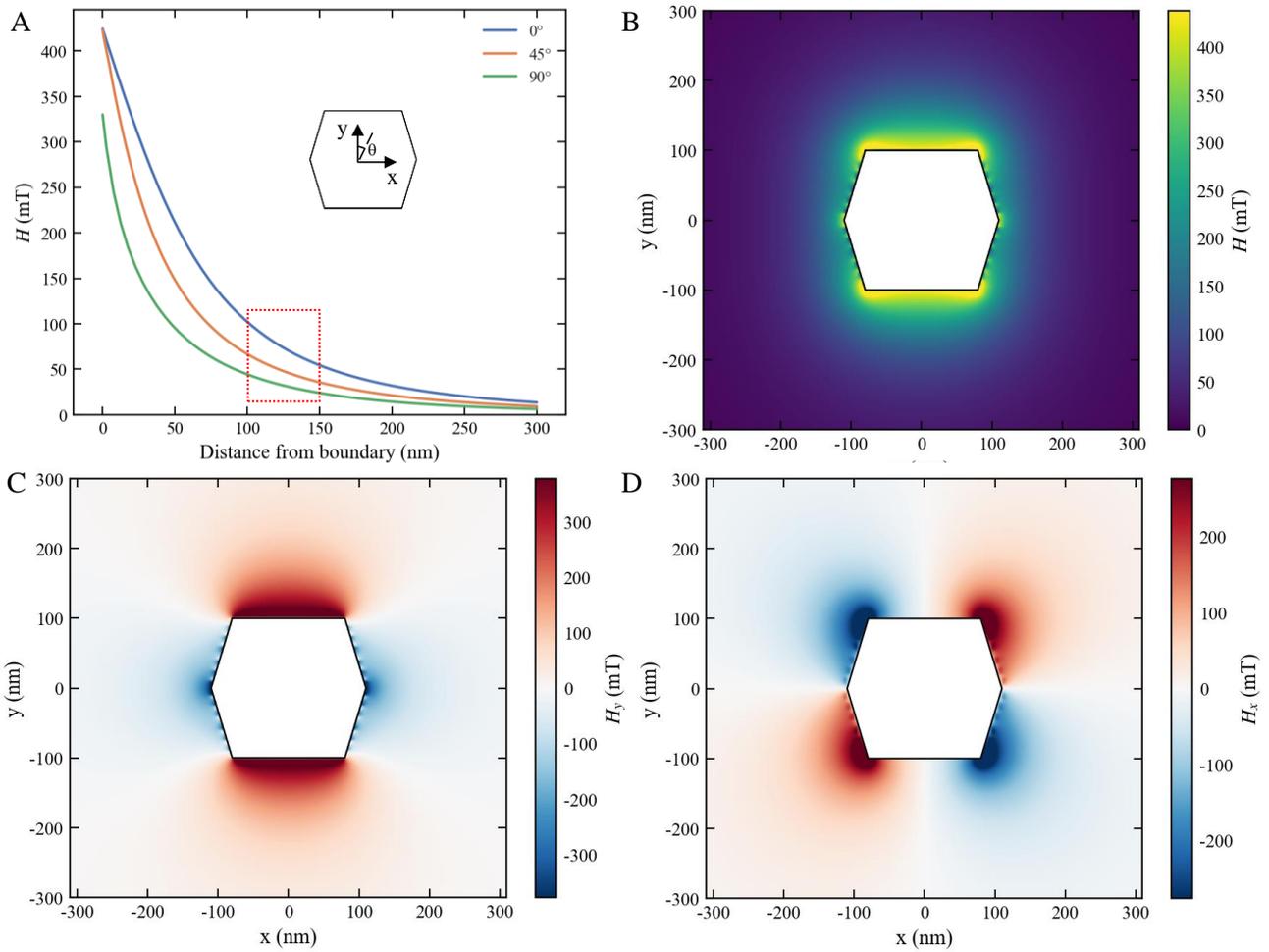

**Fig. S2 Micromagnetic calculation of dipolar fields generated by single SmCo$_5$ nanomagnet.**
(A) Dipolar fields generated by one nanomagnet versus the distance from the nanomagnet edge along different directions. The dashed box indicates that within 100-150 nm distance, the dipolar fields range from 20 mT to 100 mT. (B) Contour plot of the dipolar field strength and its component along x-axis (C) and y-axis (D) directions. These calculations assume a hexagonal nanomagnet with lateral dimension of 200 nm and thickness of 150 nm.

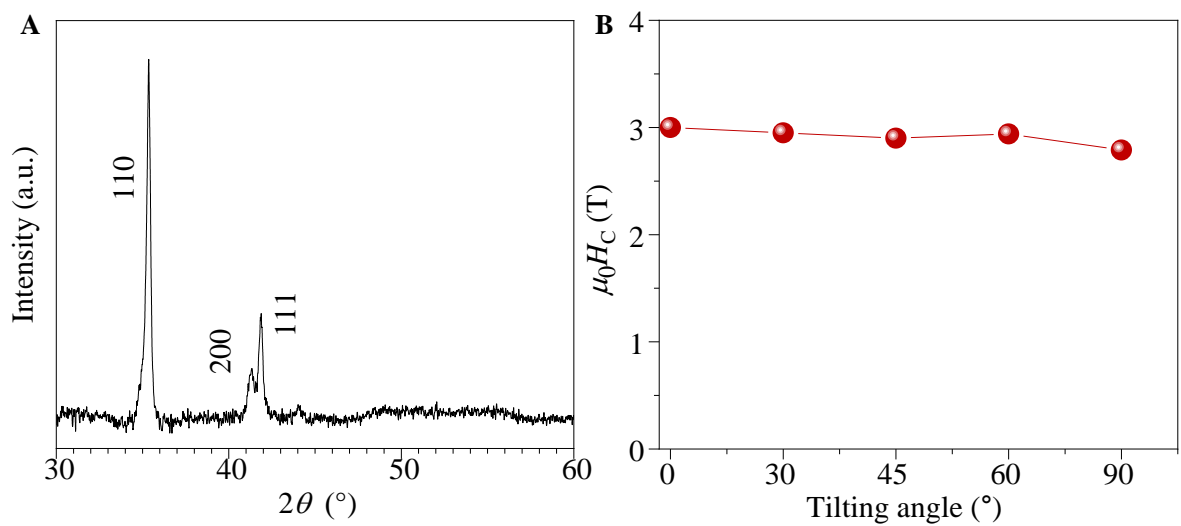

**Fig. S3** (A) X-ray diffraction pattern of the SmCo$_5$ nanomagnets network, confirming CuCa$_5$-type crystal structure with largely random crystal orientation and slight (110) texture. (B) The coercivity of the SmCo5 network in its as-prepared state versus the tilt angles from the film surface normal, indicating no pronounced angle dependence. These results indicate that, at the network level, the distribution of magnetic easy axes is effectively disordered and the macroscopic anisotropy is weak.

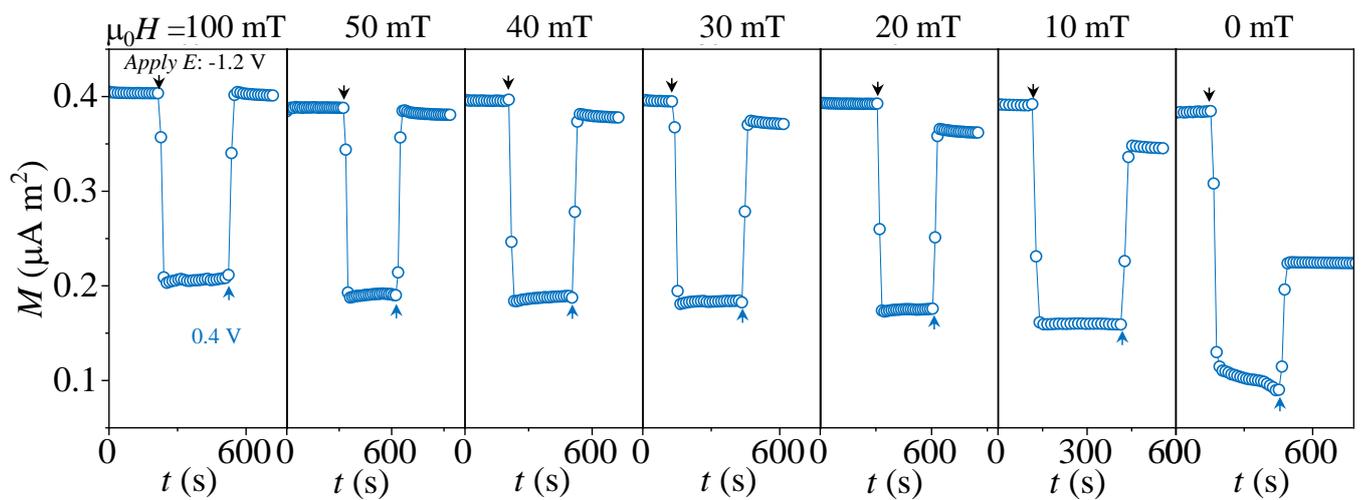

**Fig. S4** Dependence of voltage-triggered demagnetization (-1.2 V) on the external magnetic fields. Demagnetization is observed only when the external field is below ~50 mT, indicating that fields larger than ~50 mT are sufficient to overcome local dipolar fields and suppress dipolar-driven spin reversal. The initial state of the network was prepared by magnetizing at –0.4 V under 7 T, followed by reducing the field to the values indicated at the top of each plot. Black and light-blue arrows mark the times when –1.2 V and –0.4 V were applied, respectively.

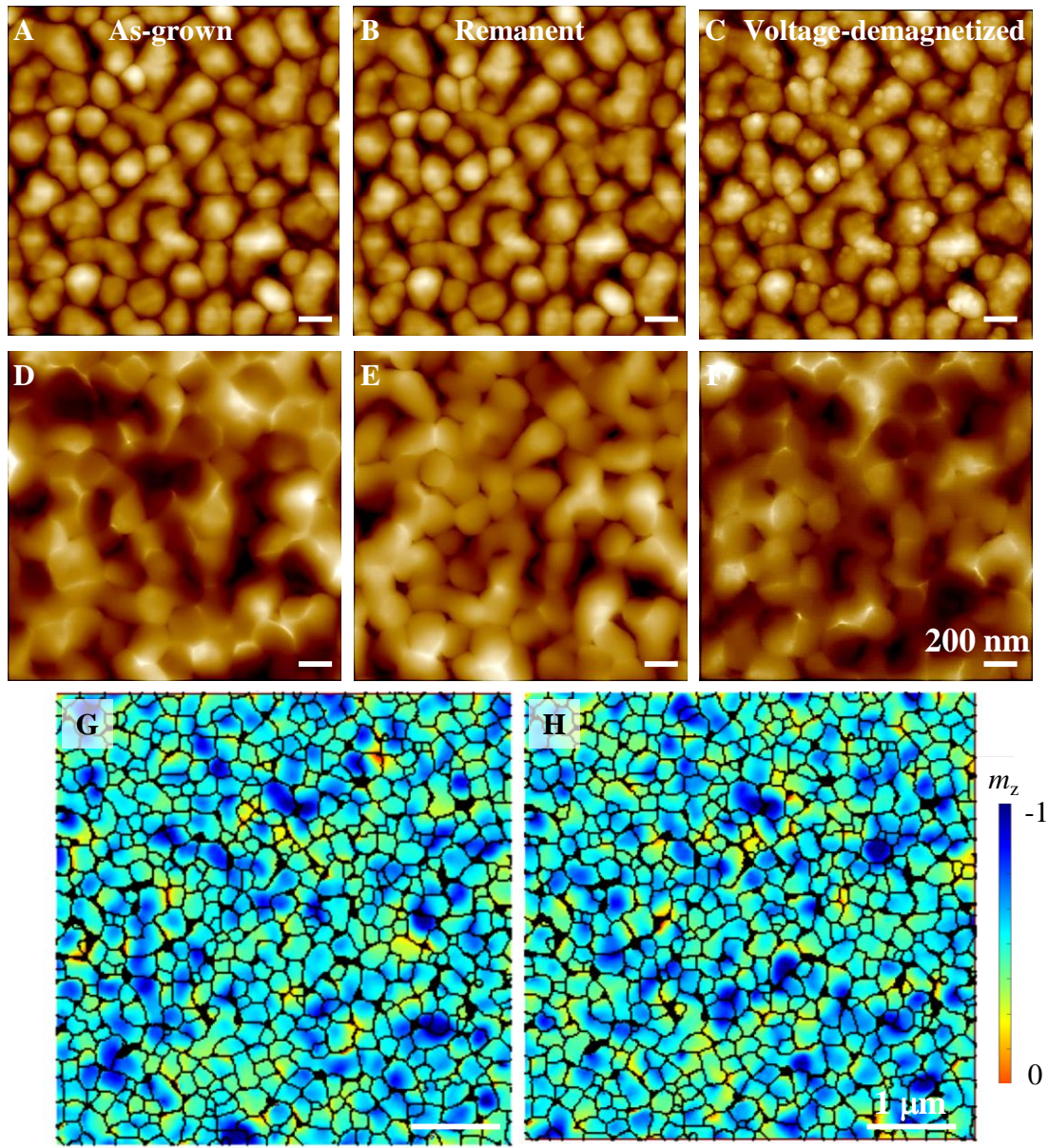

**Fig. S5 Correlated AFM and MFM mapping of the strongly dipolar-coupled SmCo$_5$ network in three distinct states.** (A, D) as-deposited (thermally-demagnetized) state, (B, E) remanent state after full saturation at 7 T and (C, F) voltage-demagnetized state after applying -1.2 V. The AFM images (A–C) confirm that the surface morphology is essentially unchanged across all states, while MFM images (D-F) reveal pronounced differences in magnetic contrast, evidencing reconfigurable domain structures governed by dipolar interactions. (G) Visualization of reversed nanomagnets in thermally and voltage-demagnetized state over larger scale (5 μm×5 μm). Images were generated by subtracting the MFM signals of the thermally-demagnetized and the voltage-demagnetized states from that of the remanent state. Blue color contrast with $m_z$ of -1 highlights nanomagnets that spontaneously flipped relative to the fully magnetized reference.

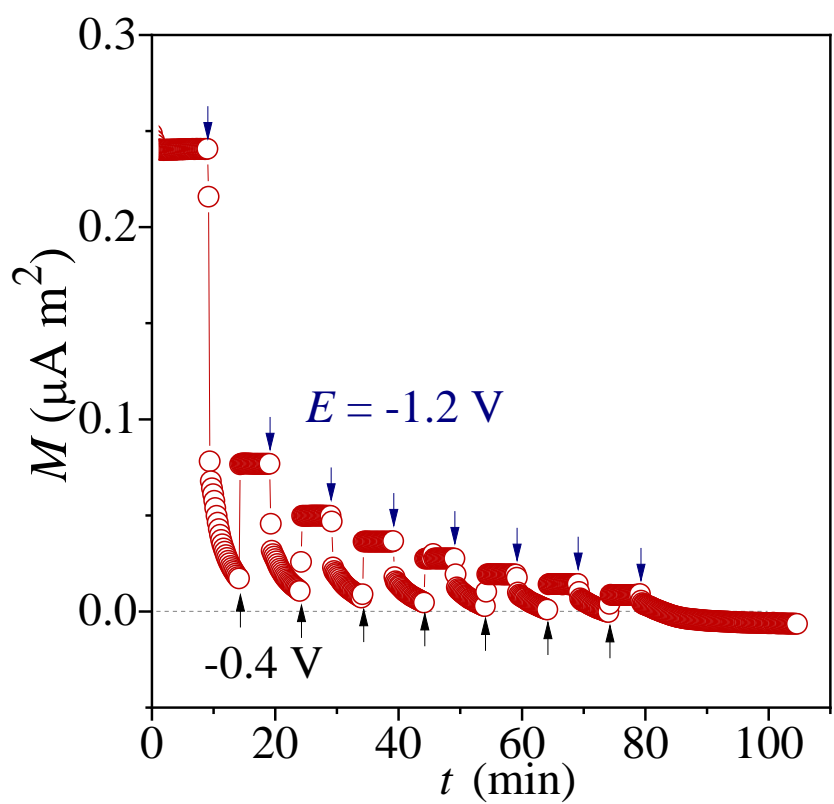

**Fig. S6 Convergent flow of magnetic states under small bias field of 3 mT.**
Convergent evolution of magnetization under a small bias field of 3 mT, following the protocol of Fig. 4A but starting from the opposite initial magnetization polarity. The net undergoes slow, stochastic spin updates over a timescale exceeding tens of minutes, converging toward an dipolar interaction-determined low-energy configuration. This long-time relaxation contrasts sharply with the fast (<10 s) voltage-induced modulation observed in Fig. 3, highlighting the difference between single-macrospin response and collective, thermally assisted dipolar dynamics at the network level.

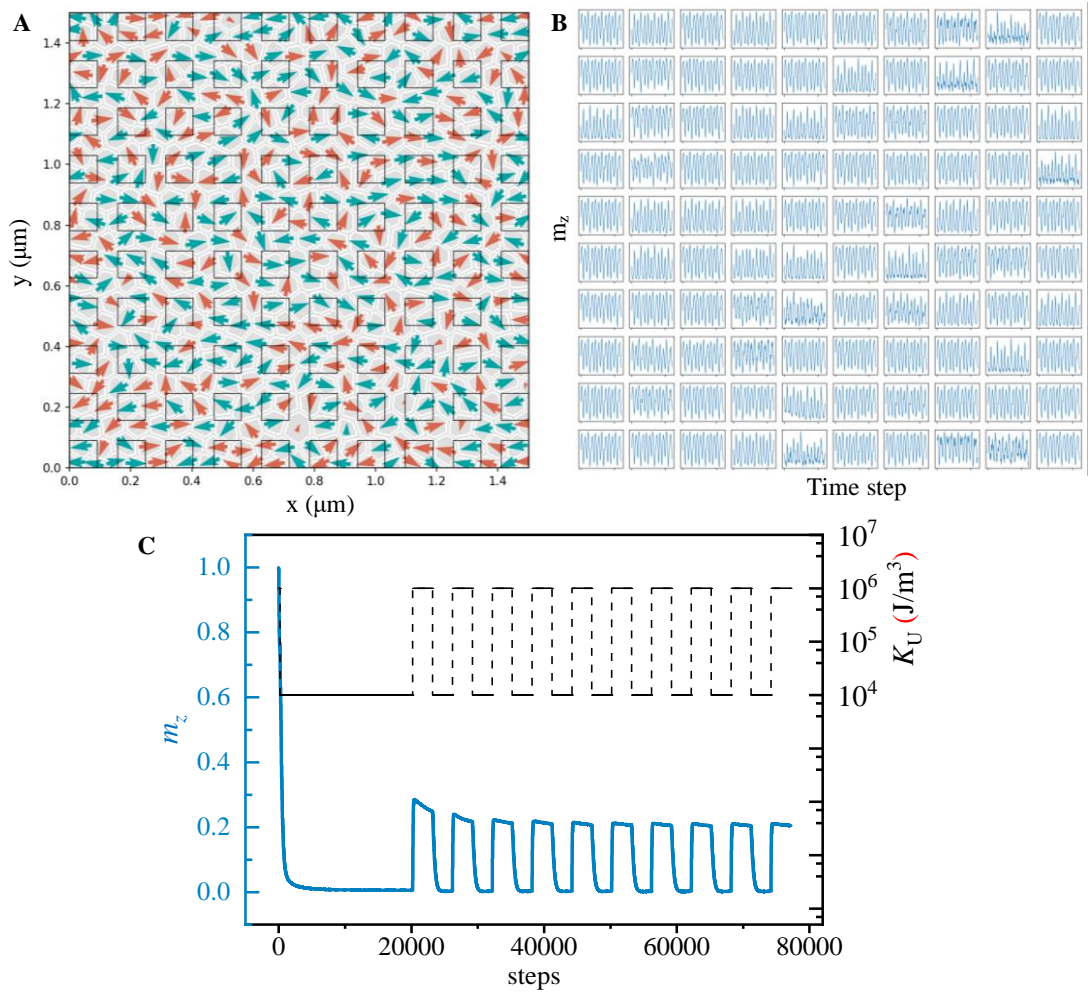

**Fig. S7 Micromagnetic construction of a strongly dipolar-coupled macrospin network and its interaction-driven dynamics.** (A) Schematic of the micromagnetic model representing a strongly dipolar-coupled network of 440 SmCo$_5$ grains (macrospins), arranged in a disordered geometry and interacting via dipolar fields. (B) Response of the global out-of-plane magnetization $m_z$ of the network to cyclic variations of the magnetocrystalline anisotropy $K_u$ between $10^6$ J/m$^3$ and $10^4$ J/m$^3$. Note that the saturation magnetization of each nanomagnet is kept fixed in these simulations, so the reversible modulation of $m_z$ arises purely from cooperative reorientation of macrospins in the demagnetized state. This reproduces the enhanced reversible modulation of remanent magnetization experimentally observed in Fig. 3A, B. (C) Dynamic evolution of the out-of-plane magnetization in 100 small blocks of the network, each block covering 2–3 macrospins, revealing reversible oscillation perturbed by the change of magnetocrystalline anisotropy. These 100 time-dependent signals constitute a high-dimensional reservoir state used for offline linear training in the reservoir-computing tasks.